\documentclass[a4paper, amsfonts, amssymb, amsmath, reprint, showkeys, nofootinbib, twoside,superscriptaddress]{revtex4-2}
\usepackage[english]{babel}
\usepackage[utf8]{inputenc}
\usepackage[colorinlistoftodos, color=green!40, prependcaption]{todonotes}
\usepackage[left=15mm,right=15mm,top=20mm,columnsep=15pt]{geometry} 
\usepackage[pdftex, pdftitle={Article}, pdfauthor={Author}]{hyperref} 
\begin{document}
\title{Towards Transcranial 3D Ultrasound Localization Microscopy of the Nonhuman Primate Brain}

\author{Paul Xing}
    \affiliation{Department of Engineering Physics, Polytechnique Montréal, Montreal, Canada}
\author{Vincent Perrot}
    \affiliation{Department of Engineering Physics, Polytechnique Montréal, Montreal, Canada}
\author{Adan Ulises Dominguez-Vargas}
\affiliation{Département de Neurosciences, Faculté de Médecine, Université de Montréal, Montreal, Canada}
\author{Stephan Quessy}
\affiliation{Département de Neurosciences, Faculté de Médecine, Université de Montréal, Montreal, Canada}
\author{Numa Dancause}
    \affiliation{Département de Neurosciences, Faculté de Médecine, Université de Montréal, Montreal, Canada}
     \affiliation{Centre interdisciplinaire de recherche sur le cerveau et l’apprentissage (CIRCA), Université de Montréal, Montreal, Canada}
\author{Jean Provost}
    \email[Correspondence email address: ]{jean.provost@polymtl.ca}
    \affiliation{Department of Engineering Physics, Polytechnique Montréal, Montreal, Canada}
     \affiliation{Montreal Heart Institute, Montreal, Canada}

\date{\today} 

\begin{abstract}
Hemodynamic changes occur in stroke and neurodegenerative diseases. Developing imaging techniques allowing the \textit{in vivo} visualization and quantification of cerebral blood flow would help better understand the underlying mechanism of those cerebrovascular diseases. 3D ultrasound localization microscopy (ULM) is a novel technology that can map the microvasculature of the brain at large depth and has been mainly used until now in rodents. Here, we demonstrated the feasibility of 3D ULM of the nonhuman primate (NHP) brain with a single 256-channels programmable ultrasound scanner. We achieved a highly resolved vascular map of the macaque brain at large depth in presence of craniotomy and durectomy using an 8-MHz multiplexed matrix probe. We were able to distinguish vessels as small as 26.9 $\mu$m. We also demonstrated that transcranial imaging of the macaque brain at similar depth was feasible using a 3-MHz probe and achieved a resolution of 60.4 $\mu$m. This work paves the way to clinical application of 3D ULM.

\end{abstract}


\maketitle

\begin{figure*}[t]
    \centering
    \includegraphics[width=0.9\linewidth]{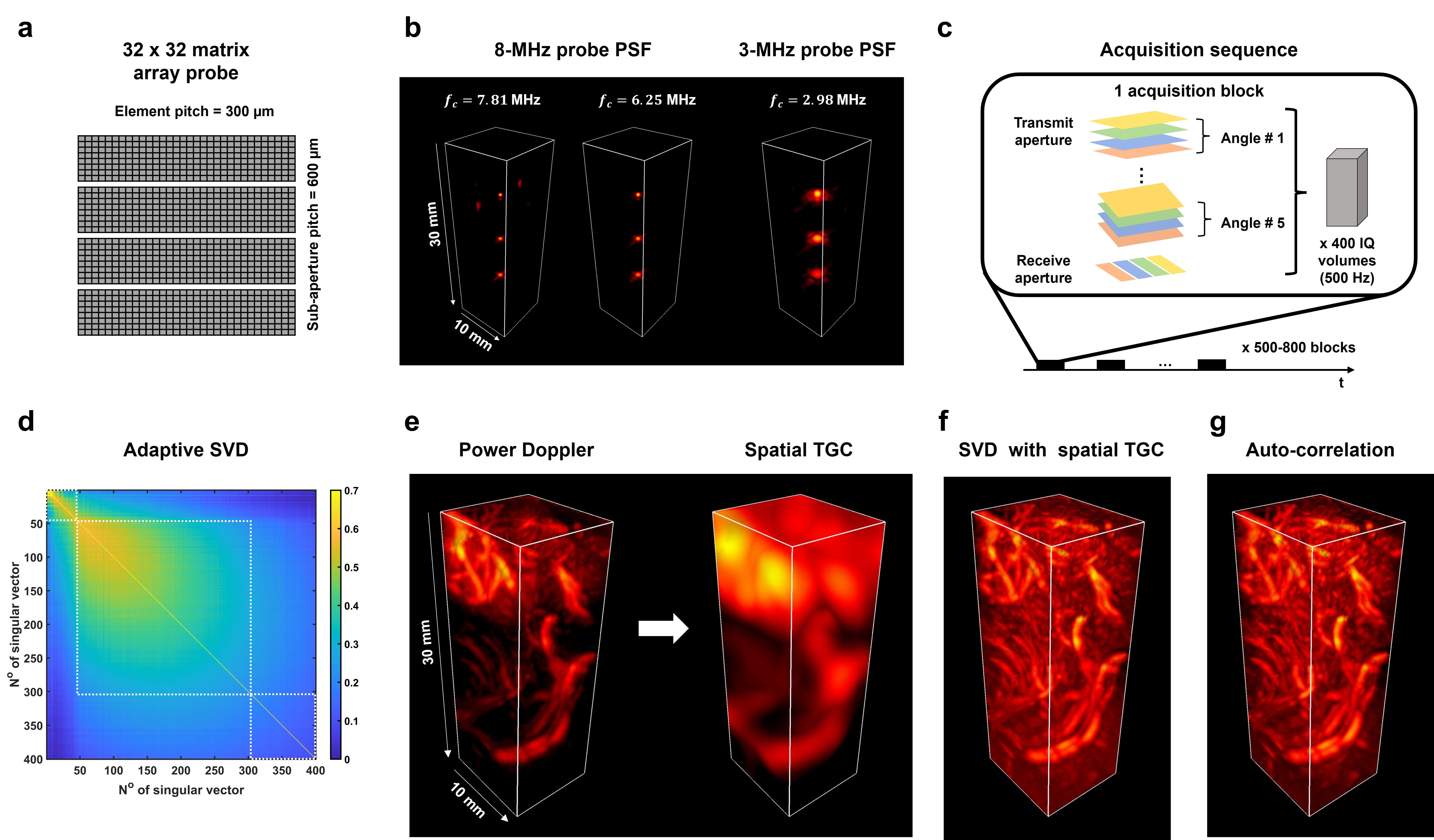}
    \caption{3D ultrafast ultrasound acquisition with a multiplexed matrix probe. (a) The $32\times32$ multiplexed matrix probe is divided into 4 active sub-apertures. (b) PSF simulation for both 8-MHz and 3-MHz probes at different depth. Grating lobes are present for pulses emitted at central frequency for the 8-MHz probe, while pulses at 6.25 MHz eliminated the grating lobes from the field of view. (c) Details of the acquisition sequence: 5 planes waves with 4 emission/reception were compounded to form a single IQ volume. Blocks of 400 IQ volumes were acquired at 500 Hz. Between 500 and 800 blocks were acquired for each experiment. (d) Adaptive SVD clutter filtering for tissue rejection. (e) Spatially dependent TGC retrieved from the Power Doppler signal after initial SVD filtering.  The Power Doppler was acquired in the presence of craniotomy and durectomy using an 8-MHz probe (f) Power Doppler with SVD filtering using the spatial TGC. (g) Power Doppler, including the spatial TGC and lag 1 auto-correlation.} 
    \label{fig:method}
\end{figure*}

\section{Introduction}

The study of the brain vasculature is of rising clinical interest, since alterations of the cerebral blood flow are present in strokes, aneurysms, and could be implicated in cognitive impairment and neurodegenerative diseases such as Alzheimer's and dementia \cite{iadecola2017neurovascular, sweeney2018role}. Developing imaging techniques for \textit{in vivo} visualization and quantification of hemodynamic changes in the brain is then key to better understand some of the underlying mechanisms of those cerebrovascular diseases and develop novel diagnostic tools and treatments in patients.

The \textit{in vivo} imaging of the brain vasculature remains challenging since it requires high penetration depth, sensitivity to a wide range of blood velocities, and high resolution to image various vessel sizes, from capillaries to larger arterial vessels. Although whole-brain hemodynamic imaging is feasible, most of the current clinical imaging modalities such as computed tomography angiography (CTA) and magnetic resonance imaging (MRI) are expensive and fail to measure properly the hemodynamic changes in smaller vessels. Optical techniques such as optical coherence tomography (OCT) offer a higher spatial resolution at the microscopic scale but are limited to a few millimeters in penetration depth \cite{ibne2017optical}. Traditional ultrasound imaging techniques such as transcranial Doppler ultrasonography (TCD) represent a widely available and non-ionizing alternative to image the brain hemodynamic but are limited in sensitivity and reliable only to rapid blood flow in larger vessels \cite{aaslid1982noninvasive}.

Ultrasound localization microscopy (ULM) \cite{desailly2013sono, viessmann2013acoustic, christensen2014vivo} is a novel imaging technique that can map the brain microvasculature \textit{in vivo} at large depth \cite{errico2015ultrafast} and at a micrometric resolution, far below the traditional diffraction limit of ultrasound \cite{couture2018ultrasound}. By detecting at a sub-pixel precision and then tracking microbubbles injected through the blood flow, ULM can achieve a resolution of the order of $\frac{1}{10}$ of the imaging wavelength. In preclinical studies, ULM has been used to evaluate structural cerebrovascular changes occurring during aging and Alzheimer's disease in rodent models \cite{lowerison2022aging, lowerison2024super}. Combined with dynamic ULM, it can be used to assert hemodynamic biomarkers such as the pulsatility index \cite{bourquin2022vivo, wiersma2022retrieving, chen2023localization} or functional activity \cite{renaudin2022functional} of the whole-brain at the microscopic scale. The clinical feasibility of transcranial ULM to assert biomarkers such as blood velocity during systole or diastole in pathological cases was also recently demonstrated in 2D with a resolution of 25 $\mu$m \cite{demene2021transcranial}.
 
Although the recent breakthroughs made by ULM offer promising clinical applications to assert cerebrovascular functions \cite{song2023super}, most of the studies were performed using 2D imaging techniques, which suffer from the inherent limitation that they can only access to vessels within the imaging plane. As a consequence, 2D ULM could also lead to bias in velocity estimations since the measurements are 3D projections into the 2D imaged plane and ignore out-of-plane movement of microbubbles and compromise the development of reliable clinical biomarkers based upon hemodynamic variations. Moreover, imaging plane selection and out-of-plane motions are additional major drawbacks of 2D ULM, leading to the need to develop volumetric ultrasound imaging techniques. Preliminary attempts to translate ULM from 2D to 3D included the use of a linear probe combined with a motorized scanning system \cite{lin20173}, but did not solve the out-of-plane issues. The development of matrix array probes and 3D beamforming strategies \cite{provost20143d} allowed to fully extend ULM to volumetric microbubble detections \cite{heiles2019ultrafast, heiles2022volumetric} and transcranial 3D ULM of the brain microvasculature was achieved in small animal models such as the rat \cite{chavignon20213d, mccall2023non} and mouse \cite{demeulenaere2022vivo, mccall2023longitudinal,bourquin2024quantitative}.

However, the difference in scale between rodent models and humans \cite{defelipe2011evolution} requires adapting multiples aspects of the ultrasound imaging and processing techniques, notably imaging at greater depth while maintaining an adequate volume acquisition rate. For non-invasive application, transcranial 3D ULM of the human brain will require using transducers with lower frequencies to image through the skull. Recently, 3D ULM was performed in larger animals such as the cat in presence of craniotomy \cite{bourquin2024quantitative, ghigo2023dynamic} and transcranially in the sheep, revealing some of the deep arterial vessels part of the circle of Willis \cite{Coudert20243D}. Nevertheless, nonhuman primates (NHPs) such as the Rhesus macaque are more suitable models for establishing  potential translation of 3D ULM to clinical applications, not only because they are similar in size to humans but also in terms of the structural organization of the brain, of cognitive and locomotor abilities \cite{cook2012nonhuman}. Indeed, NHPs have well-defined brain areas with clear homologs in humans and capacity to perform complex tasks similar to human behavior\cite{de2023transcranial}. NHPs are already in used in multitudes of ultrasound imaging studies to assert higher cognitive functions \cite{dizeux2019functional, blaize2020functional, norman2021single, griggs2023decoding}.

Herein, we demonstrate that high resolution 3D ULM reconstruction is achievable at great depth in the macaque brain when imaging with a high-frequency multiplexed probe in presence of a craniotomy. We achieved a resolution of 33.9 $\mu$m and were able to visualize the neocortex as well as deeper structures of the brains such as the striatum. We also demonstrate that 3D transcranial imaging is possible with a low-frequency multiplexed matrix array probe. We achieved a resolution as high as 60.4 $\mu$m while imaging at 3 MHz through intact skull and skin, leading to the promising translational potential of 3D ULM to clinical applications.

\begin{figure}[t]
    \centering
    \includegraphics[width=1.0\linewidth]{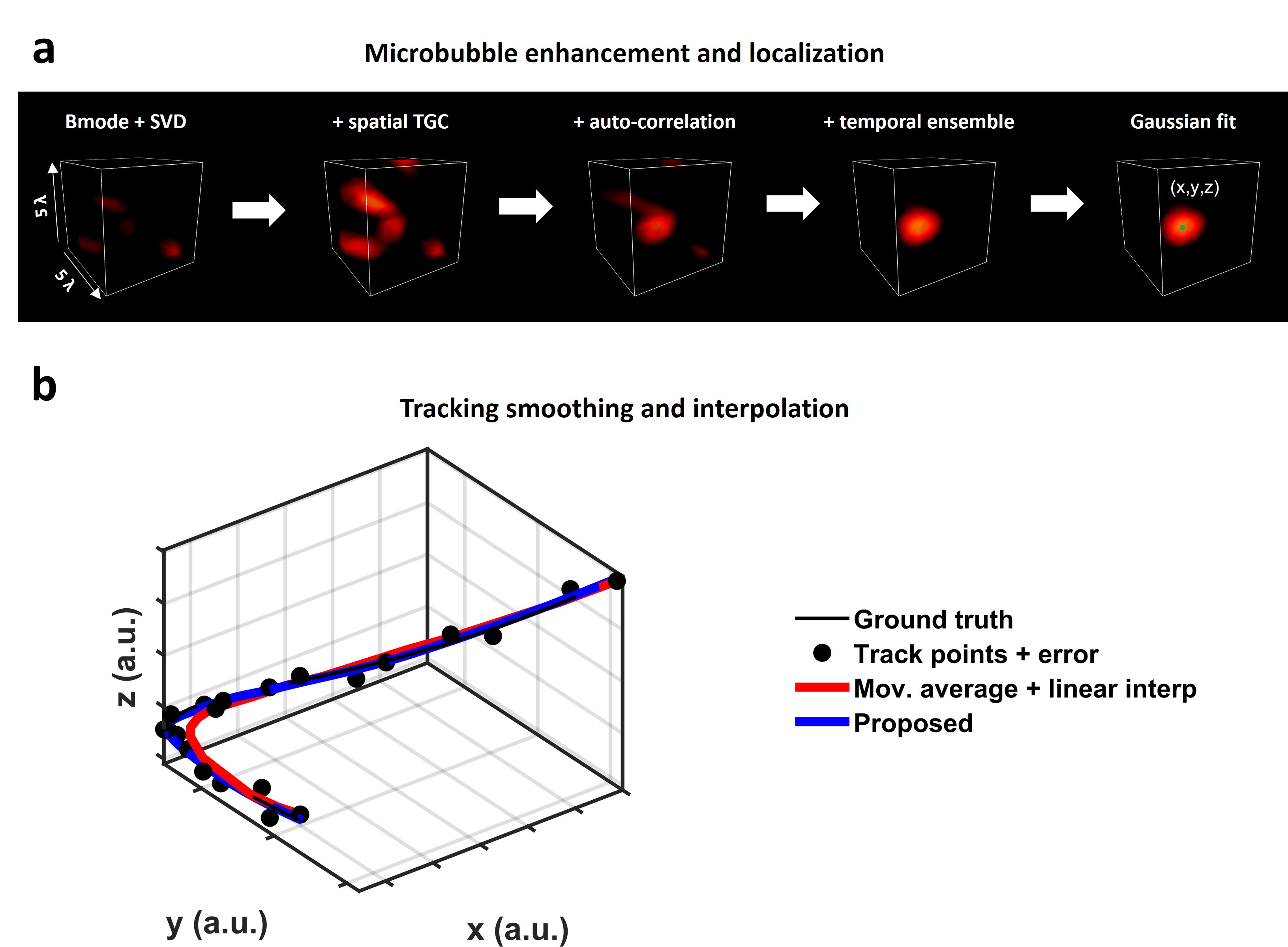}
    \caption{Microbubble localization and tracking algorithm. (a) Microbubble signals are enhanced using the spatially dependent TGC, lag 1 auto-correlation and temporal ensembling. A Gaussian fit is then used to detect the $(x,y,z)$ position of the microbubble. (b) Simulation example of the proposed smoothing and interpolation method compared to moving average and linear interpolation \cite{heiles2022performance}.}
    \label{fig:method2}
\end{figure}

\section{Methods}

\subsection{Ethics}

All experimental procedures were approved by the Animal Care Ethics Committee of the University of Montreal (Animal Care Ethics protocol numbers 20-085 and 21-065), and followed the guidelines of the Canadian Council on Animal Care, and the Animal Research: Reporting of In Vivo Experiments (ARRIVE). Two adult healthy male Rhesus macaques (\textit{Macaca mulatta}) weighing approximately 10 and 6 kg were used for this study. The 10 kg monkey was used during a terminal experiment and was previously subjected to a craniotomy on the right hemisphere exposing the primary motor cortex. The craniotomy was covered by a recording chamber that gave access to the internal capsule for the development of a localized lesion model using microstimulation. The dura mater was chronically exposed during 6 months. Macaque 2 did not undergo any medical or surgical procedure.

\subsection{Animal preparation for ultrasound imaging}

All procedures were conducted under aseptic conditions and general anesthesia. Vital signs including heart rate, respiratory rate, and blood oxygen saturation were monitored throughout the entire experiment. The body temperature was kept at 36.5-37 $^o$C using a self-regulated heating pad (Harvard Apparatus, Holliston, MA, USA). Anesthesia was induced with an injection of ketamine (10 mg/Kg, IM), given with glycopyrrolate (0.01 mg/Kg, IM) to prevent excessive salivation and maintained with 2-3$\%$ isoflurane (Furane, Baxter) in 100$\%$ oxygen.

\subsection{Experiments}
In macaque 1, the chronic chamber was first removed, and ultrasound acquisitions were performed through the chronically exposed dura mater on the ipsilesional side. After the craniotomy on the contralesional side to expose the primary motor cortex, ultrasound acquisitions were performed with and without the fresh dura mater. In macaque 2, ultrasound acquisitions were performed trough intact skull and skin. For both animals, the ultrasound probe was manually positioned perpendicularly on top of the M1 and held using a 3-axis clamp. No stereotactic reference was used for the probe placement. A distance of approximately 4 mm between the probe and the surface of imaging was kept to avoid near-field imaging.

The probe was coupled to the imaging medium with ultrasonic gel placed directly on the brain or dura mater for macaque 1, and directly on the skin for macaque 2. Definity microbubbles (Lantheus Medical Imaging, Billerica, MA, USA) were injected through a peripheral venous catheter. Bolus injections of 100-300 µL microbubbles were used combined with a 0.5 mL phosphate buffered saline (PBS) solution flush to account for the dead volume in the catheter. Additional bolus injections were performed after live monitoring confirmed elimination of previously injected microbubbles. Between 2-3 bolus injections were performed for each experiment.

\subsection{3D ultrafast ultrasound acquisitions}

Raw IQ ultrasound data were acquired with a single 256-channels Vantage system (Verasonics Inc., Redmond, WA) using the Verasonics UTA 1024-MUX adapter. A 8-MHz (7.81 MHz central frequency) multiplexed  32 $\times$ 32 matrix probe of 1024 elements (Vermon, Tours, France) was used for acquisition on macaque 1 (with craniotomy) while a similar 3-MHz (3.47 MHz central frequency) multiplexed matrix probe (Vermon, Tours, France) was used for transcranial acquisition on macaque 2. Both probes were composed of 4 sub-apertures of 32 $\times$ 8 elements with an element pitch of 300 $\mu$m $\times$ 300 $\mu$m and a sub-aperture pitch of 600 $\mu$m. The transmit frequency of the 8-MHz probe was set to 6.25 MHz to mitigate grating lobes along the sub-apertures axis (see Figure \ref{fig:method}a-b) and the transmit frequency of the 3-MHz probe was set to 2.98 MHz to minimize skull attenuation while preserving the imaging resolution.

 The ultrafast sequence (see Figure \ref{fig:method}c) was composed of 5 plane waves $\{$(-2$^o$, 0$^o$),  (-1$^o$, 0$^o$),  (0$^o$, 0$^o$),  (-1$^o$, 0$^o$),  (2$^o$, 0$^o$)$\}$ emitted with the full-aperture at the maximal pulse repetition frequency (PRF) of 16 kHz for the 8-MHz probe and at a maximal PRF of 14.7 kHz for transcranial acquisition with the 3-MHz probe, which accounts for the skull thickness. Each emission was composed of a 3-cycle pulse with 67 $\%$ duty cycle and tension of 25 V. The mechanical index (MI) was 0.05 for the 3-MHz probe and 0.07 for the 8-MHz probe. Each plane wave was acquired with four pulse emissions and received subsequently with each sub-aperture connected to 256 channels. Blocks composed of 400 volumes were acquired during 800 ms at a volume rate of 500 Hz. A total of 500 to 800 blocks were acquired for each experiment for a total acquisition time between 400 and 640 s. Volumes acquired with the 8-MHz probe were sampled with a 50 $\%$ bandwidth, while volumes acquired with the 3-MHz probe were sampled with a 100 $\%$ bandwidth.

\subsection{Spatially-dependent time gain compensation (TGC) and lag 1 auto-correlation}

\begin{figure}[t]
    \centering
    \includegraphics[width=1\linewidth]{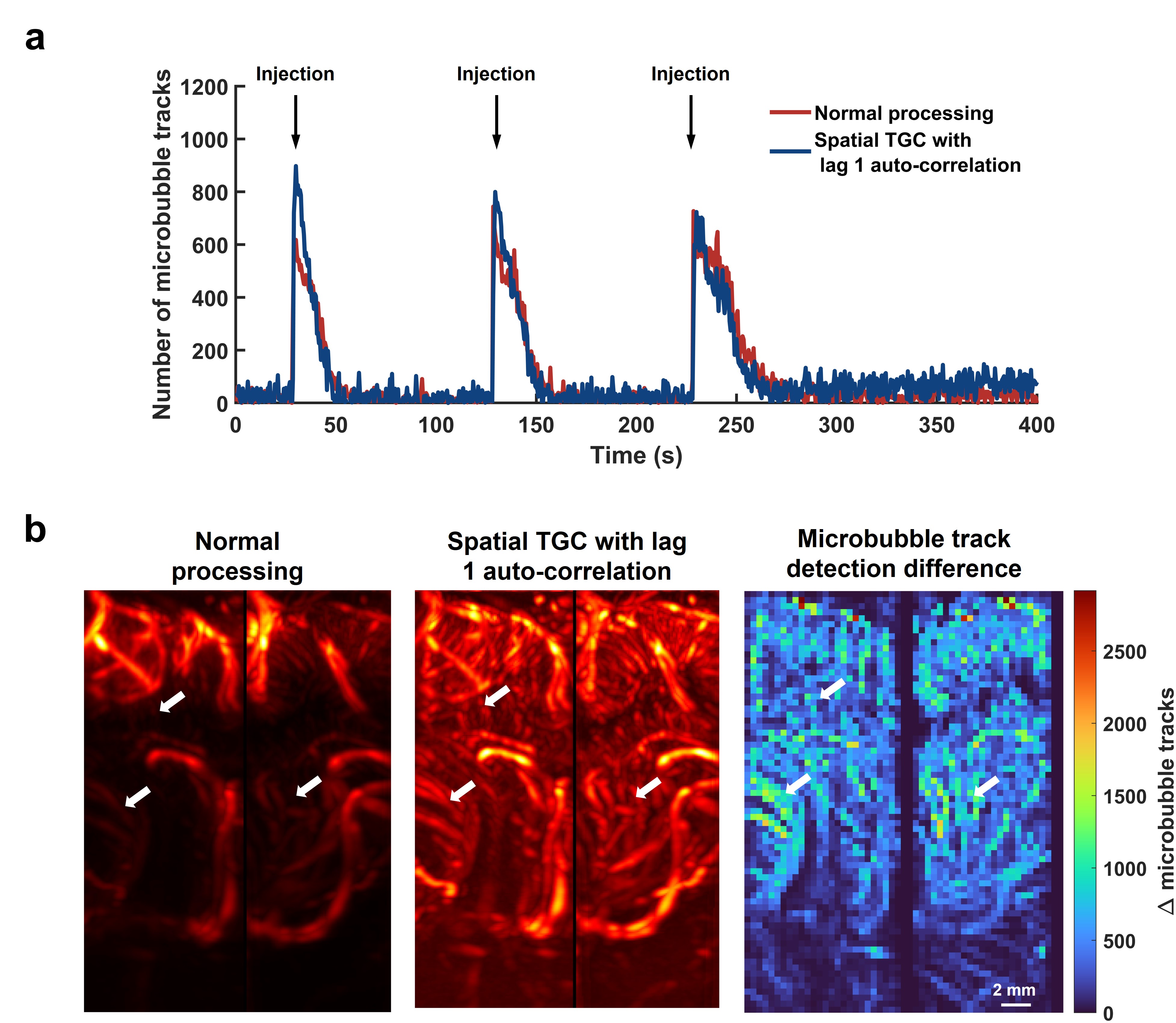}
    \caption{Impact of the spatially-dependent TGC and lag 1 auto-correlation on microbubble detection in presence of craniotomy and durectomy using an 8-MHz probe. (a) Microbubble detection counts before and after the spatial TGC and lag 1 auto-correlation. (b) Maximal intensity projections along the x and y axes of Power Doppler with the normal processing (left), Power Doppler with the spatial TGC and lag 1 auto-correlation (center) and map of the microbubble detection difference (right). Arrows indicate shadowed regions with increased microbubble track detection after spatial TGC and lag 1 auto-correlation.}
    \label{fig:figure_tgc_auto_correlation}
\end{figure}

\begin{figure*}
    \centering
    \includegraphics[width=0.9\linewidth]{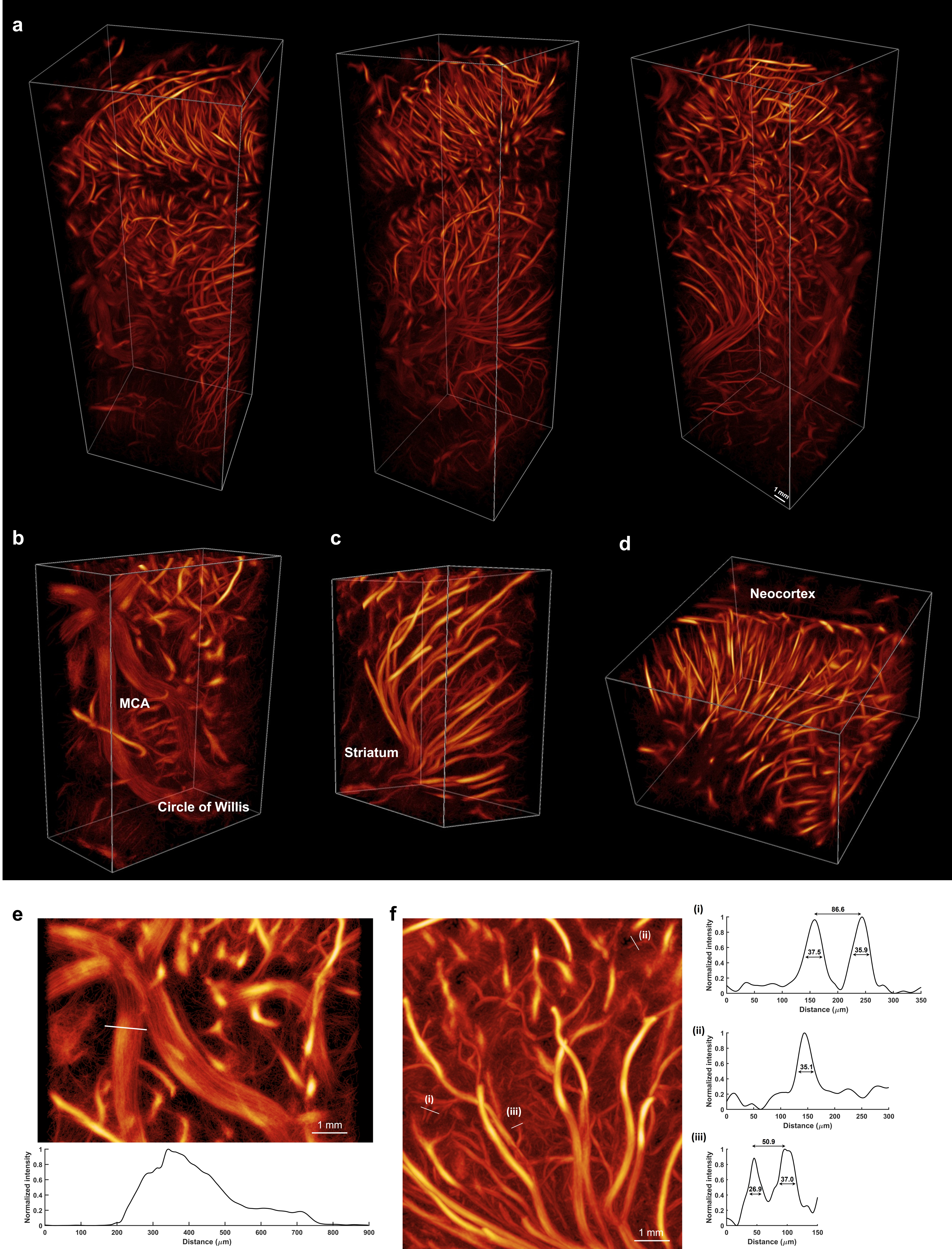}
    \caption{\textit{In vivo} 3D ULM of the macaque brain vasculature in presence of craniotomy and durectomy using an 8-MHz probe. (a) Anatomical maps of the brain vasculature with different views. (b) View on the circle of Willis and middle cerebral artery (MCA). (c) View on the striatum. (c) View on the neocortex. (e) Maximal intensity projection with profile view of a large vessel part of the MCA. (f) Maximal intensity projection with profile views of small vessels in the striatum.}
    \label{fig:ULM_craniotomy}
\end{figure*}

\begin{figure*}
    \centering
    \includegraphics[width=1\linewidth]{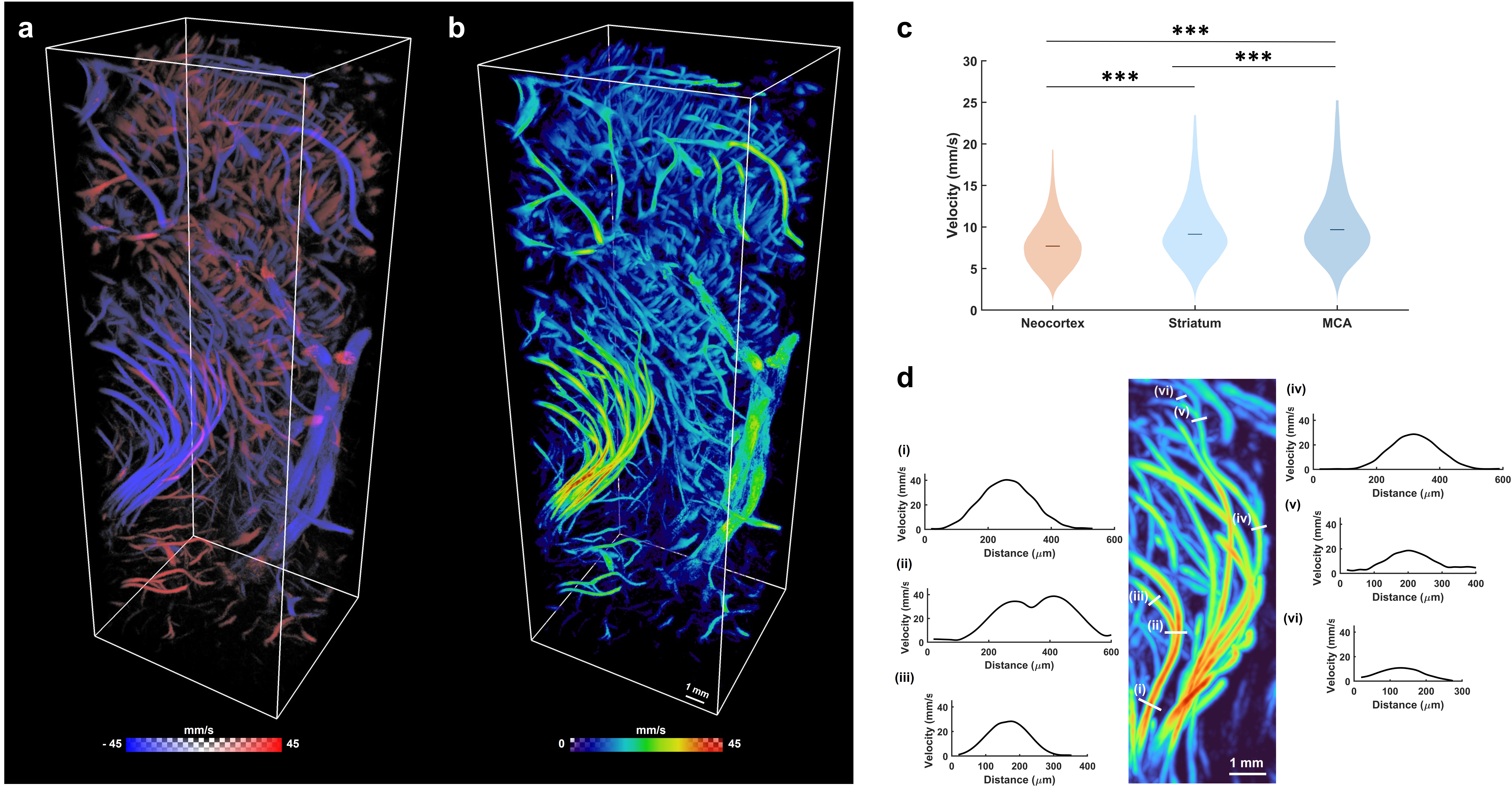}
    \caption{Hemodynamic quantification with 3D ULM of the macaque brain in presence of craniotomy and durectomy using an 8-MHz probe (a) Directional velocity map with downward flow in red and upward flow in blue. (b) Velocity magnitude map. (c) Microbubble tracks mean velocity distributions of the neocortex, striatum and middle cerebral artery (MCA). Median velocities in the neocortex, striatum and MCA are 7.7 mm/s, 9.1 mm/s, and 9.7 mm/s, respectively, and are shown with the middle bar. (d) Velocity profiles across vascular trees in the striatum. Profiles section from (i) to (iii) are part of the same vascular tree, and section from (iv) to (iv) are part of a second tree.}
    \label{fig:ULM_craniotomy_vel}
\end{figure*}

IQ data were first beamformed using a 3D delay-and-sum (DAS) algorithm based on virtual sources \cite{provost20143d} and implemented in CUDA. A $\frac{1}{2}\times\frac{1}{2}\times\frac{1}{2} \lambda^3$ grid was used. A bandpass adaptive singular value decomposition (SVD) filtering \cite{demene2015spatiotemporal, baranger2018adaptive} was performed on beamformed IQ data to remove tissue signals and noise from the microbubble signals (see figure \ref{fig:method}d). To improve signal quality in shadowed regions (see figure \ref{fig:method}e-g), a 3D Gaussian filter with standard deviation of 9 $\lambda$ was applied on the Power Doppler volumes to compute a spatially-dependent attenuation map. The beamformed volumes (before SVD filtering) were normalized with this attenuation map to alleviate shadowing artifacts. Adaptive SVD filtering were performed again on beamformed IQ data using the same threshold. A temporal lag 1 auto-correlation was then performed to further reduce noise and improve microbubble signals.

\subsection{3D ULM processing}

 Coherent temporal ensemble averaging with a Hanning window of 5 frames was performed to further enhance microbubble signals (see figure \ref{fig:method2}a). A square-root operation was applied on the auto-correlation to readjust the signal to initial dynamic range. Local maxima were detected on the envelope of the lag 1 auto-correlation by first applying a 3D dilation with a spherical kernel of radius $2.5 \lambda$. A binary mask using the correspondence of voxel intensities with the dilated volume was used to retrieve the position of the local maxima. Microbubble positions were localized at a sub-pixel resolution by using a weighted least-square Gaussian fitting \cite{guo2011simple} implemented in CUDA within a kernel size of $3\times 3\times 3 \lambda^3$. A maximum of 2048 microbubbles were detected in each volume. Only microbubbles with intensity higher than 40 dB were considered as true detections. Detected microbubbles were tracked using the Hungarian method \cite{tinevez2019simple} with no gap filling, maximal linking distance of 1 voxel (corresponding to a maximal velocity of 50 mm/s for the 8-MHz probe and 110 mm/s for the 3-MHz probe), and minimal track lengths of 10 frames. Tracks with final displacement smaller than 2 wavelengths were also rejected.

\subsection{Density mapping}

A higher dimensional smoothing algorithm based on least squares methods \cite{garcia2010robust} was applied on each track to mitigate detection errors and to retrieve the microbubble trajectory that minimize the distance between detected points (see figure \ref{fig:method2}b). The microbubble tracks were then interpolated using the modified Akima method \cite{akima1970new} to retrieve continuous trajectories while minimizing undulations. Density maps were computed by projecting and accumulating each track on a grid with voxel size of $ \frac{1}{10}\times \frac{1}{10} \times  \frac{1}{10} \lambda^3$. A spatial interpolation sampling step matching the projection grid was used to avoid gap in vessel reconstruction, and each track was only counted once by voxel \cite{heiles2022performance}. A 3D Gaussian filtering with standard deviation of 1 voxel was applied to improve 3D rendering.

\subsection{Velocity mapping}

Instantaneous velocities were calculated with a forward finite difference scheme by using the smoothed track positions. The instantaneous velocities were then interpolated using again the modified Akima method with the same sampling step used for the density map to fill the reconstruction grid after projection. Computing the velocities before interpolation ensured a more continuous profile and reduced the risk of undulations between interpolated points. After projection onto the reconstruction grid, velocities were normalized with the number of detected microbubbles \cite{heiles2022performance}. A binary mask was applied to remove velocity values in voxels where only a single microbubble was detected. A 3D Gaussian filtering with standard deviation of 1 voxel was used to further reduce noise.

\subsection{Motion correction}

Small rigid motion drifts between acquisition blocks were corrected directly on ULM map using a phase correlation algorithm \cite{hingot2017subwavelength}. To reduce sparsity, ULM blocks were binned together using a window size of 20 blocks. The position of the maximal peak of the cross-correlation between each ULM block $V_i$ and a reference block $V_{ref}$ was used to calculate spatial displacement
\begin{align}
    (\Delta x_i, \Delta y_i, \Delta z_i) = \text{argmax}_{(x,y,z)} \{\mathcal{F}^{-1}\{  \mathcal{F}\{ V_{ref}\}\cdot  \mathcal{F}\{V_i\}^*\}.
\end{align}

 Motion correction was performed for each binned block by translating the position of each voxel with respect to the calculated displacement $(\Delta x_i, \Delta y_i, \Delta z_i)$.

\subsection{Spatial resolution}

The spatial resolution was measured with the Fourier Shell Correlation (FSC) \cite{hingot2021measuring} by adapting codes from \url{https://github.com/bionanoimaging/cellSTORM-MATLAB}. After motion correction, tracks were randomly separated into two sub-images. The FSC was computed by using correlation between the spatial spectrum $F_1$ and $F_2$ of each sub-image for voxels within a radius $r$

\begin{align}
    FSC(r) = \frac{\sum_{r\in R}F_1(r)\cdot F_2(r)^*}{\sqrt{\sum_{r\in R}|F_1(r)|^2\cdot \sum_{r\in R}|F_2(r)|^2}}.
\end{align}

The intersection of the FSC curve with the half-bit threshold was used to establish the resolution. A voxel size of $ \frac{1}{20}\times \frac{1}{20} \times  \frac{1}{20}\lambda^3$ was used to ensure sufficient spatial frequency range, which also corresponds to the smaller voxel size allowed by the memory constraints.

\subsection{Statistical analysis}
All statistical analyses were performed using MATLAB. A one-way ANOVA was used to assess difference between multiple groups. A post hoc pairwise comparison using a two-sample Kolmogorov–Smirnov (KS) test was used to evaluate statistical difference between distributions. A p-value of $<0.05$ was considered statically significant. Levels of significance are given as * : p$< 0.05$, ** : p$< 0.01$, and *** : p$< 0.001$.

\section{Results}

\begin{figure}
    \centering
    \includegraphics[width=1\linewidth]{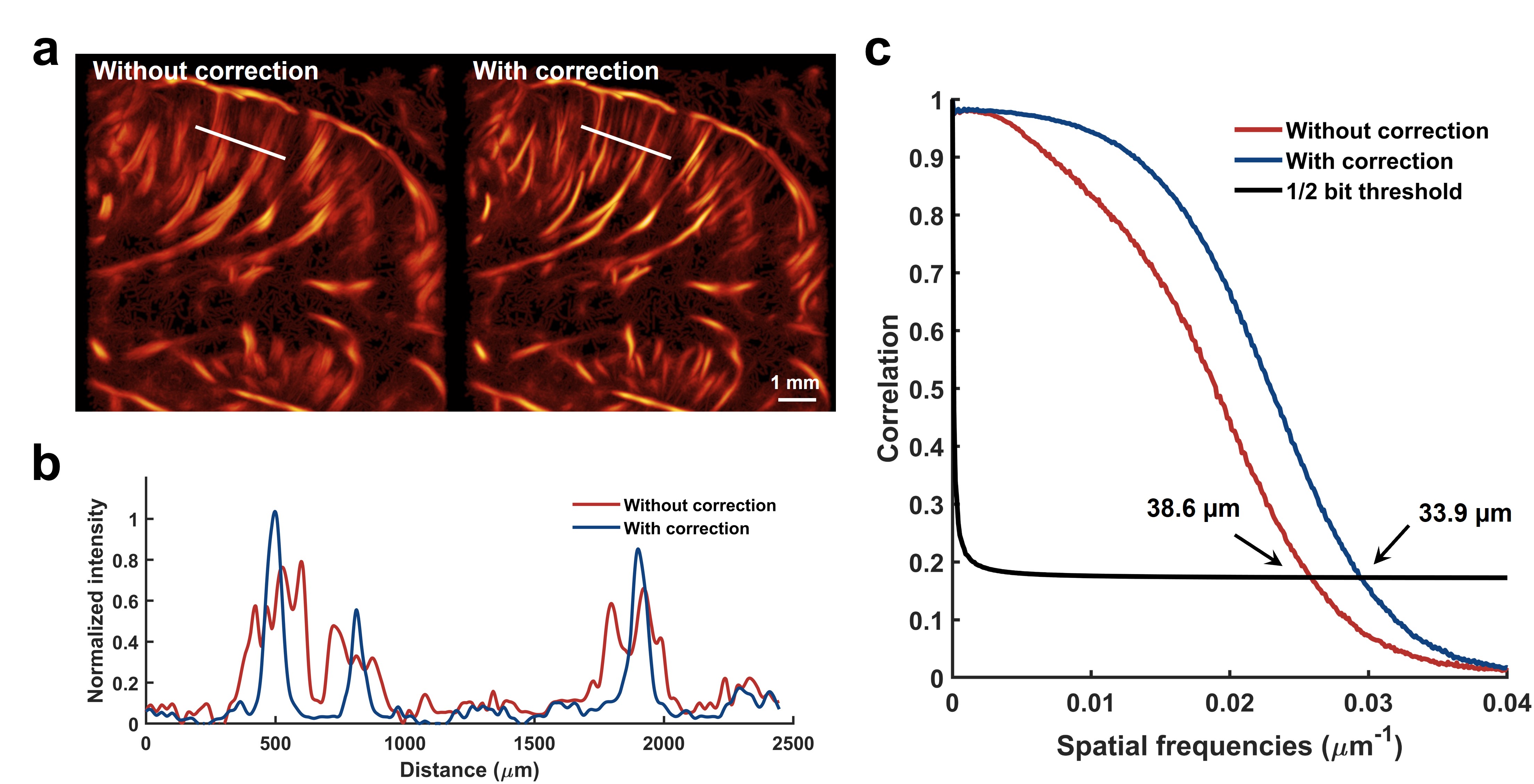}
    \caption{Effect of motion correction on 3D ULM reconstruction of the macaque brain in presence of craniotomy and durectomy. (a) Comparison of maximal intensity projection of 3D ULM without and with motion correction. (b) Profile views extracted from (a) showed that motion correction reduced duplication and increased vessel intensities. (c). Motion correction improved resolution measurement using the FSC.}
    \label{fig:motion_correction}
\end{figure}

\begin{figure*}
    \centering
        \includegraphics[width=1.0\linewidth]{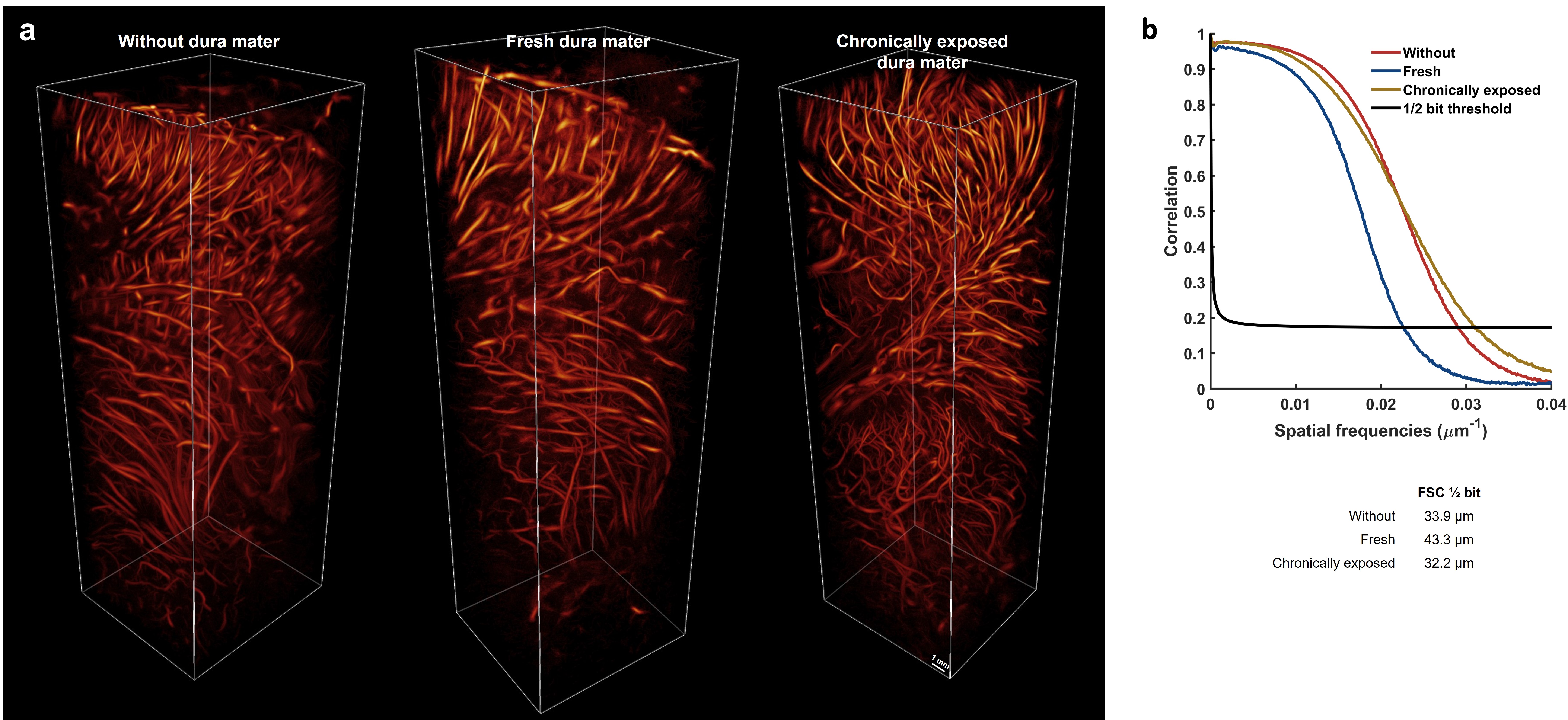}
    \caption{Impact of the dura mater on 3D ULM of the macaque brain using an 8-MHz probe. (a) Density maps without dura mater, with fresh dura mater, and chronically-exposed dura mater. (b) FSC measurements for resolution quantification.}
    \label{fig:figure_dura_mater}
\end{figure*}

\begin{figure*}
    \centering
            \includegraphics[width=1.0\linewidth]{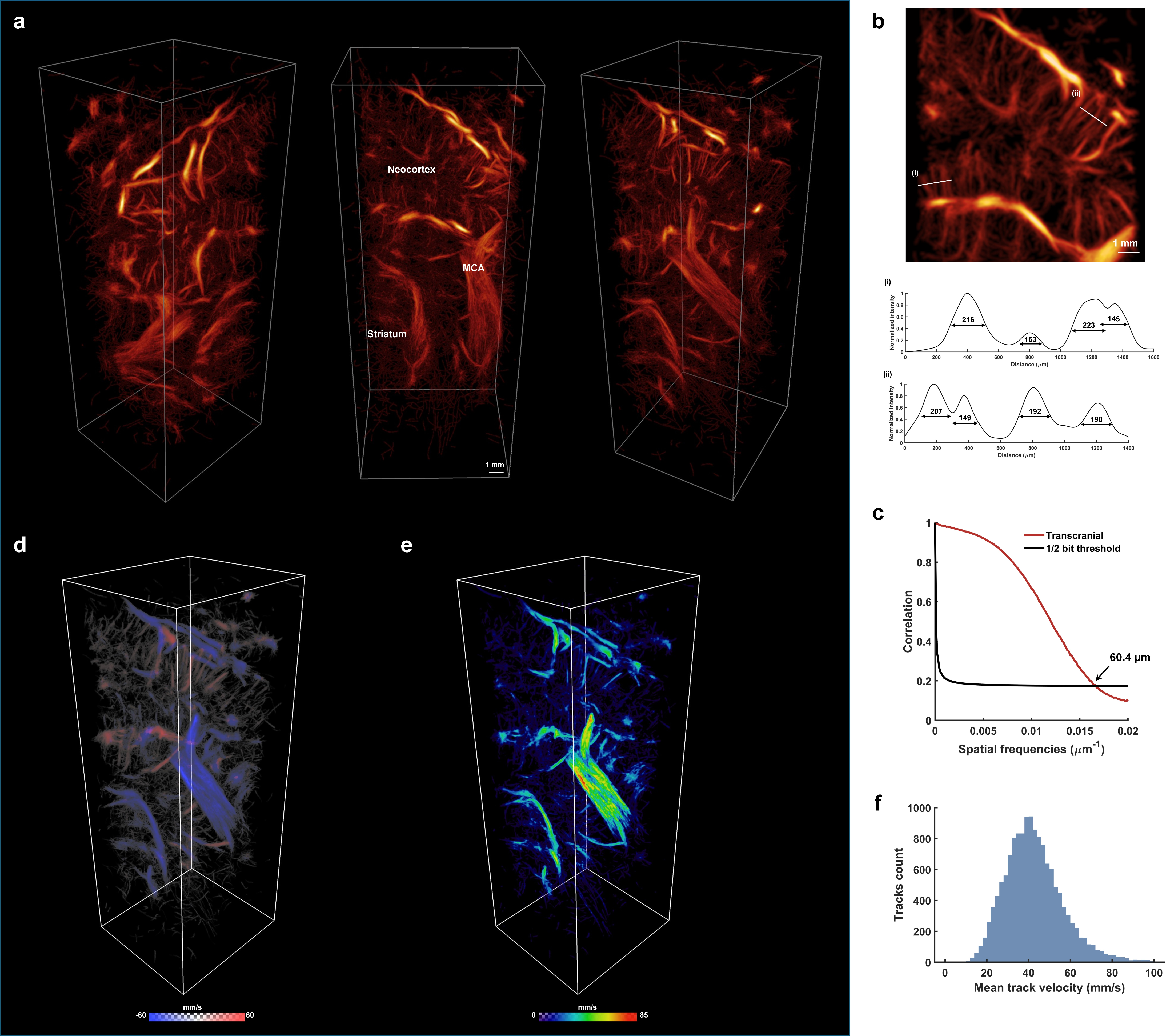}
    \caption{\textit{In vivo} transcranial 3D ULM of the macaque brain vasculature trough intact skull and skin using a 3-MHz probe. (a) Anatomical maps of the brain vasculature with different views. Anatomical regions such as the neocortex, striatum and middle cerebral artery (MCA) can be identified. (b) Maximal intensity projection with profile view of vessels in the neocortex. (c) FSC measurement established a resolution of 60.4 $\mu$m or $\lambda$/8. (d) Directional velocity map with downward flow in red and upward flow in blue. (e) Velocity magnitude map. (f) Histogram of the microbubble track mean velocities.}
    \label{fig:figure_transcranial}
\end{figure*}

\subsection{Spatially-dependent TGC and auto-correlation increase microbubble detection in shadowed regions}

Acquisition with an 8-MHz probe in presence of complete craniotomy and durectomy allowed to image deep into the macaque brain. However, some region of the brain showed important shadowing artifacts caused by the presence of large vessels. The use of a spatially-dependent TGC and lag 1 auto-correlation to alleviate shadowing artifacts increased the detection rate of microbubble tracks in each acquisition block, as shown in Figure \ref{fig:figure_tgc_auto_correlation}a. The localization heatmap in Figure \ref{fig:figure_tgc_auto_correlation}b showed the detection difference of microbubble tracks between the normal processing and processing with the use of a spatially-dependent TGC and lag 1 auto-correlation. A higher increase in microbubble tracks in shadowed region after TGC compensation are detected.

\subsection{High resolution 3D vascular map of the macaque brain with craniotomy}

In the presence of a complete craniotomy and durectomy, 3D ULM acquired with an 8-MHz probe achieved a highly resolved vascular map of the macaque brain all through the imaging depth of 3 cm (see  Figure \ref{fig:ULM_craniotomy}a), enough to image deeper structures originating from the circle of Willis such as the middle cerebral artery (MCA) as shown in Figure \ref{fig:ULM_craniotomy}b. Vascular organization of the neocortex could clearly be separated from other brain structures such as the striatum by their orientations and shapes. Indeed, vessels in the neocortex are organized in a more vertical structure, while vessels in the striatum showed a radial organization as shown in Figure \ref{fig:ULM_craniotomy}c-d. 3D ULM was able to reconstruct a wide range of vessel sizes, from large arteries like the MCA with diameter of 0.5-0.6 mm and smaller vessels with diameters down to 26.9 $\mu$m.

The directional velocity map in Figure. \ref{fig:ULM_craniotomy_vel}a showing upward and downward flows further enlightened the structural organization of the macaque brain. In the neocortex, penetrating arterioles can be separated from venules. Large upward feeding arteries originated from the base of the brain can clearly be distinguished from other structures of the brain. A wide range of velocities was also measurable with 3D ULM, ranging from 1-2 mm/s in small vessels to 45 mm/s in arteries of the circle of Willis and MCA, as shown in Figure \ref{fig:ULM_craniotomy_vel}b. Microbubble track mean velocity distribution in the neocortex was lower than in the striatum and MCA (KS test, p $<$ 0.001), while MCA microbubble tracks mean velocities were higher than the striatum (KS test, p $<$ 0.001). Figure \ref{fig:ULM_craniotomy_vel} shows the velocity profile along two different vascular trees in the striatum. Selected vessels showed a parabolic velocity profile consistent with blood flow that decreased along the vasculature.

\subsection{Motion correction improved 3D ULM resolution}

Motion correction was also a critical step to improve 3D ULM rendering of the macaque brain. Indeed, in the absence of correction, vessels appeared blurred or duplicated in the cortical region as shown in Figure \ref{fig:motion_correction}, while they appeared noticeably sharper after correction. Profile view showed that vessel are narrower and more intense with reduced replications caused by motion. Resolution measurement with the FSC established an improvement in global resolution from 38.6 $\mu$m to 33.9 $\mu$m after motion correction.

\subsection{Impact of the dura mater on 3D ULM}

A comparison of 3D ULM reconstruction imaged without the dura mater, through the fresh dura mater, and chronically exposed dura mater are presented in Figure \ref{fig:figure_dura_mater}a. Volumetric vascular maps were reconstructed at similar depth for all acquisitions. While the presence of the fresh dura mater decreased the overall imaging quality, the thickness of the chronically exposed dura mater had a limited impact. The FSC measurements presented in Figure \ref{fig:figure_dura_mater}b showed that resolution in presence of the fresh dura mater was decreased to 43.3 $\mu$m. Resolution in presence of the chronically exposed dura mater was 32.2 $\mu$m, similar to the 33.9 $\mu$m resolution measured in absence of the dura mater.

\subsection{Transcranial 3D ULM of the macaque brain with a low-frequency probe}

Figure \ref{fig:figure_transcranial} shows 3D imaging of the macaque brain through intact skull and skin using a 3-MHz multiplexed matrix probe. Volumetric ULM reconstruction of the brain vasculature at a depth of 3 cm was feasible, event in the presence of the attenuation induced by the skull. Notably, arterial branches originating from the MCA can be observed at the base of the brain as shown in Figure \ref{fig:figure_transcranial}a, and smaller vessels in the neocortex were also visible. Profile views showed that 3D ULM was capable of recontrusting vessels in the neocortex with diameters in the 100-200 $\mu$m range, as shown in Figure \ref{fig:figure_transcranial}b. The FSC measurement established a global resolution of 60.4 $\mu$m, or an 8-fold improvement over the diffraction limit of resolution. The directional velocity map in Figure \ref{fig:figure_transcranial}d showed the presence of penetrating arterioles and venules in the neocortex that can be distinguished from their downward or upward flow, respectively. The hemodynamic quantification presented in Figure \ref{fig:figure_transcranial}e-f showed that cerebral blood velocities ranged from less than 5 mm/s in small vessels to 85 m/s in larger branches of the MCA with a microbubble track mean velocity distribution between 10-100 mm/s.

\section{Discussion}

In this work, we demonstrated the feasibility of \textit{in vivo} 3D ULM of the NHP brain at large depth using a multiplexed 1024-element matrix probe driven by a single 256-channel Verasonic system. We introduced novel processing techniques to improve signal quality of microbubbles and alleviate shadowing artifacts caused by large vessels. We evaluated the impact of the different layers surrounding the brain of the Rhesus macaque on imaging quality, from the dura mater to the skull. We achieved a resolution of 33.9 $\mu$m in presence of a craniotomy and durectomy while using an 8-MHz probe and a resolution of 60.4 $\mu$m transcranially trough intact skull and skin using a 3-MHz probe. This represents an 8-fold increase in resolution over traditional diffraction-limited ultrasound imaging techniques.

In the presence of craniotomy and durectomy, 3D ULM achieved a highly resolved reconstruction of the microvasculature of the Rhesus macaque neocortex as well as of deeper structures of the brain. We were able to image down to 3 cm, which represents a 3-4 fold increase in imaging depth compared to previous studies performed in rodents \cite{chavignon20213d, demeulenaere2022vivo}. Large arterial structures including the MCA could be reconstructed, as well as vessels as small as 26.9 $\mu$m in the striatal region. The dense and rich vascular organization of penetrating arterioles and venules of the neocortex of the Rhesus macaque could be observed. By analyzing vessel orientations, 3D ULM was also able to identify clear anatomical regions of the brain, such as the striatum.

The imaging technique also achieves visualization of blood velocities from a large dynamic range, with higher velocities up to 45 mm/s in the larger ascending arteries part of the circle of Willis such as branches of the MCA, as well a slower flow down to 1-2 mm/s in smaller vessels in the neocortex. 3D ULM could provide hemodynamic quantification throughout the imaging depth, and could distinguish arterioles from venules in the neocortex from their flow direction. The propagation of blood flow dynamics along the vascular tree in deeper regions of the brain could be observed, such as the decrease of the velocity along the upstream and downstream arterial branches of the striatum.

The dura mater, the layer of conjunctive tissue surrounding the brain, can reach a thickness of 0.5 cm in primates\cite{kinaci2020histologic, galashan2011new}, which is comparable to the skull thickness in rodents \cite{o2011ultrasound} and could have limited ultrasound transmission at high frequency. We were still able to image at a depth of 3 cm in presence of the dura mater with an 8-MHz probe and measure similar resolution around of 32-34 $\mu$m using the FSC between the chronically exposed dura mater and without dura mater acquisitions. Interestingly, the acquisition with the fresh dura mater had the lowest resolution, with an FSC measurement of 43.3 $\mu$m. Acute swelling and more prominent non-rigid brain motion directly following the craniotomy could have impaired the quality of the acquisition on the fresh dura mater. Since the probe was manually placed without stereotactic references, the exact same imaging volume could not be retrieved between experiments. The volume acquired through the exposed dura mater was also performed on the ipsilesional side (i.e., the side of the original lesion), which could further account for anatomical difference and could have affected the FSC measurements. Nevertheless, those findings support that imaging through the dura mater is feasible with an 8-MHz probe even when the dura mater was thicker after being chronically exposed.

By improving microbubble signal through auto-correlation and temporal ensembling to compensate for skull attenuation, transcranial imaging with intact skull and skin was also feasible with a 3-MHz probe with similar depth. The transcranial acquisition was performed on top of the frontal/parietal bone of the Rhesus macaque which, with a thickness up to 4 mm \cite{adams2007biocompatible}, is similar to the thickness of the human temporal bone \cite{kwon2006thickness} used as an imaging window in clinical settings. Hence, the results presented in Figure \ref{fig:figure_transcranial} could be representative of what could be achieved non-invasively in humans with 3D ULM using lower frequency probes. Moreover, anatomical regions could also be clearly identified, such as the large vessels originating from the MCA and the striatal arteries, but also some of the penetrating arterioles and venules in the neocortex. The lower resolution of transcranial 3D ULM can be explained by the lower frequency used for imaging, the low sensitivity of the probe, as well by the absence of aberration correction method which could have impaired microbubble detection. The sequence was also not fully optimized for transcranial imaging, which could have improved further the microbubble signals. The microbubble PSF also appears larger at lower frequencies. Imaging volume filling occurs then at lower microbubble concentrations, requiring longer acquisition time to form a complete ULM map \cite{belgharbi2023anatomically}. Continuous intravenous injection \cite{renaudin2022functional} instead of bolus injections of microbubbles could have help improve ULM reconstruction when longer acquisitions are required.

This work paves the way for both intraoperative and non-invasive clinical applications of 3D ULM. Studies of stroke and neurodegenerative diseases could benefit from this technique. The feasibility of longitudinal studies using 3D ULM has been recently demonstrated in mice \cite{mccall2023longitudinal}. 3D ULM could offer both structural and hemodynamic insights \textit{in vivo} at the microscopic scale on disease progression, such as vascular remodeling after brain injuries or neurovascular impairment in Alzheimer's disease. 2D transcranial ULM has been used to measure hemodynamic changes such as systole and diastole blood velocities to characterize pathological cases such as brain aneurysm in patients \cite{demene2021transcranial}. The extension to 3D will offer more robust biomarkers based on velocity measurements. Dynamic ULM has been recently extended to 3D \cite{bourquin2024quantitative, ghigo2023dynamic} and could help further establish new hemodynamic biomarkers for neurovascular diseases by offering temporal information through the entire cardiac cycle. 

Although this study serves as a proof-of-concept of the potential feasibility of transcranial 3D ULM in humans using a multiplexed matrix probe, it still suffers from certain limitations. Only two Rhesus macaques were used, which limits the validation and reproducibility of the results. While NHPs access are limited and therefore cannot be used in high numbers, they play a critical role in biomedical research, being more similar to humans both anatomically and cognitively, and offer a unique opportunity to better understand how the human brain works.

Vessel reconstructions and appearances in ULM can be highly dependent on the tracking algorithm used and the number of detected microbubbles travelling through the lumen of each vessel. Larger vessels require a higher number of microbubble accumulation to be completely spatially sampled \cite{hingot2019microvascular}, which could explain why part of the circle of Willis and MCA appeared less pronounced than other regions of the brain on the anatomical map in Figure \ref{fig:ULM_craniotomy}. Microbubbles with higher velocities and larger numbers of neighbor microbubbles can also be harder to track, which could have further degraded the appearance of larger vessels. 

Moreover, the maximal linking distance used in the tracking algorithm set a global maximal velocity. The smaller linking distance used for acquisitions performed at 8-MHz improved performance locally in regions with slower flow such as the neocortex at the expense of the rendering of faster vessels such as the MCA as shown in Figure \ref{fig:ULM_craniotomy}. The transcranial acquisition at 3-MHz was able to perform better reconstruction in larger vessels since the tracking algorithm allowed a higher maximal velocity, but did not perform as well in the neocortex where velocities are lower. This could explain some of the discrepancies in velocity measurement between the 8-MHz and 3-MHz probes in the MCA and other regions with higher velocities. Developing alternative and adaptive tracking algorithms such as Kalman filtering \cite{tang2020kalman, taghavi2022ultrasound, lok2022three} could improve 3D ULM performance over a larger range of microbubble flow dynamics and increase accuracy of velocity measurements.

Skull-induced aberration represents one of the main limitations for transcranial brain ultrasound imaging. The development of 3D aberration correction methods \cite{bureau2023three} compatible with ultrafast ultrasound imaging, like those already used for 2D ULM \cite{demene2021transcranial, robin2023vivo, xing2023phase, xing2024inverse} could significantly improve image formation by increasing microbubble detection and reducing potential vessel duplication artifacts. 

The low-frequency matrix probe combined with plane waves transmission has a limited field of view not compatible for whole-brain imaging, especially in humans. The use of matrix probes combined with divergent wave transmission \cite{provost20143d} like used in echocardiography \cite{papadacci20204d} or cylindrical wave like recently used in the sheep \cite{Coudert20243D} could help extend the field of view. The development of novel matrix probes with larger elements could further help to allow whole-brain imaging \cite{favre2022boosting, favre2023transcranial}. Row–Column Array (RCA) strategies for 3D ULM \cite{jensen2019three, taghavi2022vivo, kim2022deep} could also offer promising solutions by extending the field of view while using a limited number of channels to address the elements. However, RCA probes suffer from the same limitation as multiplexed probes of a reduced volume rate (at the expense of the imaging depth) arising from the number of transmit angles necessary to satisfy the sampling requirement \cite{sauvage2018large}. 

Another drawback of 3D ULM is the large amount of data and acquisition time required to accumulate enough microbubbles in order to form a complete vascular map. Extension of novel processing techniques \cite{huang2020short, you2022curvelet, leconte2023tracking} and deep learning strategies \cite{milecki2021deep,shin2023context, rauby2024pruning} to 3D could help reduce acquisition time by increasing the microbubble detection rate and tracking performance at higher concentration. More efficient hard drives could help have faster transfer system to facilitate continuous block acquisitions. Recent research in novel probe designs with fewer elements could help reduced hardware complexity and data management by spatially encoding the received signals and moving the complexity to the image reconstruction algorithm end \cite{caron2023ergodic, brown2024four}. Current processing time also limits the real-time application of 3D ULM. Furthermore, the continuing increasing computational power and the development of highly paralleled systems could help significantly reduced processing time in the future to facilitate clinical application of 3D ULM.

\section{Conclusion}

In this work, we report the feasibility of 3D ULM of the NHP brain by using a multiplexed matrix array probe and a single 256-channels Verasonics Vantage system. In the presence of a craniotomy and durectomy, we could image at a depth of 3 cm and achieved a resolution of 33.9 $\mu$m using an 8-MHz probe. We also demonstrated that transcranial 3D ULM is feasible at similar depth using a 3-MHz probe and achieved a resolution of 60.4 $\mu$m.

\section{Acknowledgments}
This work was supported in part by the New Frontier in Research Fund (NFRFE-2022-00590), in part by the Canada Foundation for Innovation under grant 38095, in part by the Natural Sciences and Engineering Research Council of Canada (NSERC) under discovery grant RGPIN-2020-06786, in part by Brain Canada under grant PSG2019, and in part by the Canadian Institutes of Health Research (CIHR) under grant  PJT-156047 and MPI-452530. Computing support was provided by the Digital Research Alliance of Canada. The work of Paul Xing was supported in part by the TransMedTech Institute and in part by NSERC.

\bibliography{references}  

\end{document}